\documentclass[twocolumn,aps,prb,superscriptaddress,floatfix,nofootinbib]{revtex4-2}
\usepackage{amsmath}
\usepackage{amsfonts}
\usepackage{amssymb}
\usepackage{graphicx}
\usepackage{color}
\usepackage{bm}
\usepackage{comment}
\usepackage[final]{hyperref} % adds hyper links inside the generated pdf file
\hypersetup{
	colorlinks=true,       % false: boxed links; true: colored links
	linkcolor=blue,        % color of internal links
	citecolor=blue,        % color of links to bibliography
	filecolor=magenta,     % color of file links
	urlcolor=blue         
}

\begin{document}

\title{Transport theory for a generic two-arm co-propagating Majorana interferometer with Majorana fermion and edge vortex tunneling}

\author{Domenico Giuliano}
\affiliation{Institut f\"ur Theoretische Physik, Heinrich-Heine-Universit\"at, D-40225 D\"usseldorf, Germany}
\affiliation{Dipartimento di Fisica, and INFN, Gruppo Collegato di Cosenza, Universit{\`a} della Calabria, Arcavacata di Rende, I-87036 Cosenza, Italy}

\author{Andrea Nava}
\affiliation{Institut f\"ur Theoretische Physik,
Heinrich-Heine-Universit\"at, D-40225  D\"usseldorf, Germany}

\author{Fabian Hassler}
\affiliation{Institute for Quantum Information, RWTH Aachen University, D-52056 Aachen, Germany}

\author{Reinhold Egger}
\affiliation{Institut f\"ur Theoretische Physik,
Heinrich-Heine-Universit\"at, D-40225  D\"usseldorf, Germany}

\begin{abstract}
    Co-propagating chiral Majorana edge modes are predicted to exist in proximity-coupled quantum anomalous Hall systems. They combine the exciting prospects of edge vortices as flying non-Abelian Ising anyons, with their fully electrical detection through anyon fusion processes. In this setup, electrical transport arises from the interference of a pair of Majorana edge states. By bosonizing the model with an additional replica sector, we generalize previous treatments to fully asymmetric configurations. We analyze a Mach-Zehnder interferometer and show that while the DC conductance does not exhibit signatures of edge-vortex interference, at low frequencies, this interference manifests as an effective capacitance. This quantity is sensitive to the non-Abelian anyon statistics of edge vortices and, in particular, carries signatures of their topological spin and nontrivial conformal dimension. We demonstrate that the thermal length $l_T = \hbar v/(\pi k_B T)$ is the relevant scale at low temperatures $T$, analogous to quasiparticle transport in conventional quantum Hall edge states with velocity $v$. As long as the interferometer arm lengths match within this length scale, transport signatures of the non-Abelian statistics of edge vortices remain visible. 
\end{abstract}
\maketitle

\section{Introduction} \label{sec1}

Anyonic excitations exhibit fractional statistics as has been  experimentally verified by measurements of the statistical braiding phase of Abelian anyons in fractional 
quantum Hall  (FQH) devices \cite{bartolomei2020,nakamura2020}.  The elusive non-Abelian statistics of the more 
complicated non-Abelian anyons predicted, e.g., for filling factor $\nu=5/2$ \cite{nayak2008}, is much harder to 
verify and is a key objective of current research.  The simplest type of non-Abelian anyon is the Ising anyon, i.e., a 
vortex harboring a Majorana zero mode (MZM) excitation.  Such excitations, once under good control, could be
very useful for topological quantum information processing applications \cite{nayak2008}.  Apart from FQH devices, 
they are also expected to arise in topological superconductors (SCs), and we focus on this platform to illustrate
our ideas below.  In two spatial dimensions, the edge modes of the simplest topological SCs are neutral
one-dimensional (1D) chiral Majorana fermion modes.  While  Majorana fermion edge modes are just fermions with a self-adjointness constraint, Ising anyons here correspond to so-called edge vortices (EVs), which are intrinsically mobile (``flying'') Ising anyons co-moving with the chiral Majorana fermions.  As such, they could serve for transporting quantum information \cite{beenakker2020}.  In the context of this paper, this aspect 
offers a convenient route to probe the effects of anyon statistics on the charge conductance of a Majorana interferometer if EV tunneling processes can occur.  It also makes the SC case simpler than the FQH case because of the absence of charged edge modes and because of the fact that we can have co-propagating edges on different sides of an interferometer.  Flying Ising anyons are closely related to spatially localized MZMs \cite{barkeshli2019} that are more widely studied. In particular, signatures of the latter  have been reported in recent experiments on proximitized semiconductor heterostructures and/or tunable quantum dot arrays, see, e.g., Refs.~\cite{dvir2023,q2023,tenHaaf2024,aghaee2025,vanLoo2026,q2026,zatelli2026}.  
Nonetheless, experimental demonstrations of the braiding of localized MZMs have not yet been reported (apart from quantum simulations \cite{stenger2021,harle2023}) and face complications due to disorder effects where conventional fermionic Andreev states can mimic many Majorana features \cite{prada2019}.  In contrast, flying anyons exploit the chirality of edge modes of a gapped topologically ordered bulk phase, where anyon braiding processes are insensitive to the effects of weak disorder \cite{bartolomei2020,nakamura2020,nayak2008,bonderson2006,bonderson2007,fendley2009,beenakker2020}. 

Probably the simplest realization of flying Ising anyons is given by EVs in chiral Majorana fermion edge states \cite{fendley2007,beenakker2020}. EVs are composite objects built of a coreless vortex (domain wall) in the chiral Majorana mode, with one-half of a fermionic flux quantum, which has a zero-energy MZM as a bound state \cite{beenakker2020}.
The domain wall corresponds to a sign change, i.e., a $\pi$-phase shift at the EV position, for the Majorana mode.  
It is worth noting that EVs only bind a single MZM but no additional finite-energy bound states (as is the case for standard vortices in SCs), and thus they are pure Ising anyons.
One may inject classical (deterministic) EVs into 1D chiral Majorana modes by applying suitable voltage pulses \cite{beenakker2019b,beenakker2019a,adagideli2020,hassler2020,flor2023}. Here, we study quantum EVs
which can be dynamically nucleated, e.g., at Josephson junctions \cite{nava2024,giuliano2026}. 
The dynamics of EVs $(\sigma)$ shows non-Abelian statistics with nontrivial correlation functions and several anyon fusion channels \cite{nilsson2010,clarke2010,hou2011,grosfeld2011,ariad2017}. To give an example, 
the long-distance equal-time correlator of two EVs is \cite{bonderson2007,fendley2007,ariad2017}
\begin{equation}\label{correl}
    \langle \sigma(x)\sigma(0) \rangle\propto e^{2\pi i s_\sigma} |x|^{-2h_\sigma}, \quad s_\sigma=h_\sigma=\frac{1}{16},
\end{equation}
where the topological spin is defined as $e^{2\pi i s_\sigma}=e^{i\pi/8}$ and the conformal scaling dimension $h_\sigma$
coincides with $s_\sigma$.  The topological spin is closely tied to the non-Abelian exchange statistics of Ising anyons, and measuring its value is tantamount to detecting non-Abelian braiding \cite{kitaev2006,bonderson2007,nava2024}.

Interferometric setups can give access to transport observables encoding the above physics.  In this work, we focus 
on the Mach-Zehnder interferometer as the simplest type of Majorana interferometer based on two chiral, co-propagating Majorana edge modes.
This geometry has been studied theoretically in the absence of Josephson junctions inside the interferometer, and thus without Ising anyons, for topological insulator (TI) surfaces partially proximitized by a grounded SC (defining the interior region of the interferometer) and  by ferromagnets with different magnetization directions (defining the outer regions) \cite{fu2009,akhmerov2009,nilsson2010,clarke2010,hou2011,roising2018,shapiro2021,huang2026}.
Since SCs and magnets induce different gap types in the TI surface, where fermionic quasiparticles correspond to a single Dirac cone, 1D chiral Majorana edge modes must exist at interfaces between gapped regions of different (SC vs magnetic) type \cite{hasan2010}.
Even though clear experimental evidence for chiral Majorana modes is not yet available, proximity-induced SC in quantum anomalous Hall insulators has been reported in Refs.~\cite{amet2016,lee2017,uday2024}.
Importantly, by choosing opposite magnetization directions of the ferromagnets, both chiral Majorana edge modes co-propagate along the same direction, in contrast to FQH interferometers \cite{nayak2008,beenakker2020}.  This fact leads to important simplifications below.  

\begin{figure}[tb]
    \centering
    \includegraphics[width=\linewidth]{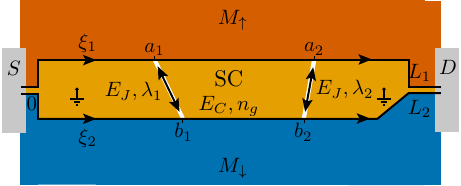} 
    \caption{Schematic setup for an asymmetric chiral Majorana interferometer formed on the surface of a TI.  In the interferometer region, the surface is proximitized by a conventional SC.  Outer regions are proximitized by magnetic materials of opposite magnetization  ($M_\uparrow$ and $M_\downarrow$). At interfaces between magnetic materials, 1D chiral Dirac fermion channels (double lines) arise.
    Arrows denote the propagation direction. At interfaces between SCs and magnets, we have a pair of chiral 1D Majorana fermion modes $\xi_{1,2}(x)$  (black lines) with different length, $0<x<L_{1,2}$.   These modes are populated from the 1D Dirac channel emanating from the normal source electrode $S$. Similarly, outgoing particles are absorbed by the normal drain electrode $D$.    
    Because of chirality and the existence of a bulk gap, the conversion between Dirac and Majorana modes occurs with unit probability \cite{fu2009,akhmerov2009}. A central floating SC island with charging energy $E_C$ and offset charge parameter $n_g$ is formed by two Josephson line junctions (with Josephson energy $E_J$). The first (second) line junction connects the upper Majorana mode at $x=a_1$ ($x=a_2$) and the lower mode at $x=b_1$ ($x=b_2$), which triggers Majorana fermion and EV tunneling processes (indicated by the double-arrowed lines).  The outer SC regions are grounded.  For  details, see Sec.~\ref{sec2}. 
    }
    \label{fig1}
\end{figure}

By locally connecting both arms of the above interferometer \cite{fu2009,akhmerov2009} via Josephson line junctions within the SC region, Majorana fermions can tunnel between the chiral Majorana edge modes.
In addition, quantum EVs are nucleated at the line junctions \cite{akhmerov2009,nava2024,giuliano2026}.
The two line junctions here define a central floating island with a charging energy $E_C$, where we consider
the regime $E_C\ll E_J$ with the Josephson energy $E_J$ of the line junctions.  (The outer SC regions of the device 
remain grounded.)
In Ref.~\cite{nava2024}, we have analyzed a highly symmetric geometry of the device in Fig.~\ref{fig1}, where the two Majorana edge modes had the same length, $L_1=L_2$, and the tunneling points on both arms were chosen identical, $a_1=b_1$ and $a_2=b_2$ in Fig.~\ref{fig1}.
The purpose of the present paper is to generalize the theory presented in Ref.~\cite{nava2024} to allow for 
arbitrary length mismatch $\delta L=L_1-L_2$ and arbitrary tunneling points with $0<a_1<a_2<L_1$ and $0<b_1<b_2<L_2$.  As we will show below, this generalization 
is by no means trivial and requires conceptual leaps since the technical toolbox used in Ref.~\cite{nava2024} is no longer applicable.

As shown in Ref.~\cite{nava2024} for the symmetric device, the low-frequency AC charge conductance $G(\omega)$, in particular the effective capacitance $C_{\rm eff}$, 
of the device in Fig.~\ref{fig1} encodes  information about the topological spin of EVs, and thus about anyon statistics.
Such measurements are expected to be much easier than the shot noise and/or collision experiments 
proposed for revealing non-Abelian braiding statistics in FQH devices \cite{lee2022}, see also Ref.~\cite{Beenakker2025}, since conductance measurements refer to an averaged quantity instead of subtle correlation functions, see also Refs.~\cite{tang2026,dewit2026}.  
(However, see Ref.~\cite{alkalay2026} for a very recent experimental report of such an experiment.)
For the symmetric geometry considered in Ref.~\cite{nava2024}, the DC conductance $G(\omega=0)$ does not reveal any signature of anyon statistics because of the simplicity of interfering trajectories.  We note that in multiterminal
geometries, the DC conductance is already expected to reveal anyon statistics \cite{giuliano2026}.  An interesting question addressed below is  whether generic asymmetric two-terminal devices as shown in Fig.~\ref{fig1} will contain signatures of 
anyon statistics in $G(0)$.  In addition, using the theory presented below, a quantitative theory of the effective capacitance $C_{\rm eff}$ will be given free from geometric restrictions.

The remainder of this paper has the following structure. In Sec.~\ref{sec2}, we describe the device in Fig.~\ref{fig1}
in detail, and start to formulate the theory without EV tunneling in terms of a fermionic description.
In Sec.~\ref{sec3}, we introduce a replicated bosonization field theory in order to tackle the inclusion of 
EV tunneling processes and formulate the full low-energy theory for asymmetric geometries.
In Sec.~\ref{sec4}, we then apply our theory to calculate the linear-response conductance tensor. 
In Sec.~\ref{sec5}, we summarize our results and offer an outlook.  
We mostly put $\hbar=k_B=1$ in what follows. 

\section{Model} \label{sec2}

The device in Fig.~\ref{fig1} features two chiral Majorana fermion modes of length $L_1$ and $L_2$, respectively, 
which can be realized on a TI surface that is proximitized inside the interferometer by a grounded SC with fixed phase, say, $\varphi=0$ \cite{fu2009,akhmerov2009}. 
A 3D TI material such as Bi$_2$Se$_3$ has a single 2D gapless Dirac fermion cone at its surface, which may 
be gapped out by the proximity effect using either a SC or a magnet \cite{hasan2010}. In practice, one deposits either 
a conventional SC film or a ferromagnetic film on the surface.  
At the interfaces between different-type gapped regions, 1D chiral Dirac or Majorana edge modes are present.
For simplicity, we assume that all modes have the same edge velocity $v$. The spatial coordinate
$0\le x\le L_{1,2}$ is used separately for the two chiral Majorana modes.
Outside the interferometer region, the TI surface is instead proximitized by ferromagnets with opposite magnetization, 
with 1D chiral Dirac fermion modes at their interfaces.
We assume that the induced SC pairing gap $\Delta$ is below the corresponding magnetic gaps, such that
the only available low-energy modes below $\Delta$ are the neutral 1D chiral Majorana modes and the charged 1D chiral Dirac fermion modes \cite{fu2009,akhmerov2009}. The energy scale $\Delta$ thus also acts as the bandwidth of the low-energy theory below.  Note that the grounded SC regions ensure charge conservation at 
the  points where Dirac fermions are converted into pairs of Majorana fermions, see Fig.~\ref{fig1}.
Due to the chirality and the existence of a bulk gap, these Dirac-Majorana conversion processes happen with unit probability \cite{fu2009,akhmerov2009}.  

We next assume that in the grounded SC regions in Fig.~\ref{fig1}, the strips have a sufficiently large width, $W\gg l_c= v/\Delta$, such that co-propagating chiral Majorana modes at opposite edges have  negligibly small overlap except at the Josephson line junctions.  Similarly, we require $L_{1,2}\gg l_c$.  Using typical values in proximitized TI devices, say, $v\sim 10^4$~m/s 
and $\Delta/h\sim 100$~GHz \cite{hasan2010}, the typical coherence length is $l_c=v/\Delta\sim 15$~nm.  
Particles are then injected into the chiral Majorana modes via a 1D chiral Dirac channel from the normal source electrode $S$ with chemical potential $\mu_S(t)=eV(t)=eV\cos(\omega t)$, where we consider both the DC
regime $\omega\to 0$ and the low-frequency AC regime. Similarly, outgoing particles are absorbed by the normal drain electrode $D$ held at chemical potential $\mu_D=0$.
One then measures the current $\langle I_{D}(t) \rangle$ 
flowing through the interferometer into the drain electrode.

The device in Fig.~\ref{fig1} also features a central floating SC island with charging energy $E_C$ and  dynamical (fluctuating) phase $\varphi$.  In general, the asymmetric form of this island
is defined by the position of the two Josephson line junctions,
see Fig.~\ref{fig1}.  
We assume the same Josephson energy $E_J$ for both junctions, with $E_J\gg E_C$.  In this regime, fast and rare quantum phase slips, $\varphi\to \varphi\pm 2\pi$,
simultaneously affect both Josephson line junctions and cause composite 
EV tunneling processes.  Moreover,
the floating central SC island is capacitively coupled to a gate electrode which allows one to tune the 
offset charge $n_g$ (in units of $2e$).

Under these conditions,
important low-energy degrees of freedom in Fig.~\ref{fig1} are the 1D chiral Majorana fermion modes $\xi_{1,2}(x)$,
which are self-adjoint field operators, $\xi^\dagger_{j}(x)=\xi_{j}^{}(x)$, subject to fermionic anticommutation relations,
\begin{equation}\label{MajoAlgebra}
    \{ \xi_{j} (x), \xi_{j'}(x')\} = 2\delta_{j,j'}\delta(x-x').
\end{equation}
Both the upper ($\xi_1$) and lower ($\xi_2$) arm are assumed to have the same edge velocity $v$, but their length
can differ.  We use coordinates $0\le x\le L_{1,2}$ for the respective chiral Majorana mode.  
Without Majorana and/or EV tunneling processes, the low-energy Hamiltonian for chiral Majorana modes is given by \cite{gogolin1998}
\begin{equation}\label{MajoranaHam0}
    H_0 = - \frac{iv}{2} \sum_{j=1,2} \int_0^{L_j} d x \, \xi_{j}(x) \partial_x \xi_{j}(x)  .
\end{equation}
The normal source electrode $S$ is located at $x<-L_0$, with the Dirac-Majorana conversion happening at $x=0$,
i.e., $L_0$ denotes the length of the incoming 1D chiral Dirac channel in Fig.~\ref{fig1}.  
The corresponding Dirac fermions are described by chiral fermionic field operators $\Psi_{S}(x)$ with $-L_0<x<0$, with the Hamiltonian
\begin{equation}\label{HSD}
    H_{S} = -iv\int_{-L_0}^0 dx\, \Psi^\dagger_{S}(x)\partial_x \Psi^{}_{S} (x).
\end{equation}
After moving through the interferometer, the Majorana modes are converted back to the outgoing 1D chiral Dirac channel for which we use new coordinates with $0<\tilde x<L_0$, such that for $\tilde x>L_0$ the drain electrode $D$ absorbs all particles. 
Defining $H_D$ for the corresponding chiral Dirac field $\Psi_D(\tilde x)$ with $0<\tilde x<L_0$ in analogy to Eq.~\eqref{HSD}, the Dirac-Majorana conversion processes at $x=0$ and $\tilde x=0$ are described by the matching conditions \cite{fu2009,akhmerov2009,nilsson2010}
\begin{align}\label{match1}
\xi_{1}(0)&=\Psi_{S}^{}(0)+\Psi^\dagger_{S}(0),\,\, \xi_{2}(0)= i ( \Psi_{S}^{\dagger}(0)-\Psi^{}_{S}(0)
),\\  
\xi_{1}(L_1)&=\Psi_{D}^{}(0)+\Psi^\dagger_{D}(0),\,\, \xi_{2}(L_2)= 
i ( \Psi_{D}^{\dagger}(0)-\Psi^{}_{D}(0)).\nonumber
\end{align}

We next include Majorana fermion tunneling processes connecting both Majorana edges at the two Josephson line junctions
in Fig.~\ref{fig1}.  Note that the left (right) line junction connects the point $x=a_1$ ($x=a_2$) on the upper Majorana arm with $x=b_1$ ($x=b_2$) on the lower Majorana arm, where we always assume $0<a_1<a_2<L_1$ and $0<b_1<b_2<L_2$. 
With real-valued dimensionless Majorana tunneling amplitudes $\lambda_{1}$ and $\lambda_2$,
the standard Majorana tunneling Hamiltonian \cite{gogolin1998} is given by
\begin{equation}
 H_{\rm MT} = - iv \sum_{j=1,2}\lambda_j \xi_{1}(a_j) \xi_{2} (b_j).
\label{MTferm}
\end{equation}
Postponing the discussion of EV tunneling processes to Sec.~\ref{sec3}, we can study charge transport through the interferometer using the standard Landauer-B\"uttiker scattering matrix approach 
adapted to chiral Majorana modes \cite{li2012,strubi2011}. In fact, the
Hamiltonian $H_0+H_S+H_D+H_{\rm MT}$ together with the matching conditions \eqref{match1} is quadratic in the fermion operators.  By diagonalizing the scattering problem, we find the lengthy but exact finite-temperature ($T$) solution for the current-voltage characteristics in the absence of EV processes.  We here consider only the DC limit. With $\delta L=L_1-L_2$ and $d_j = a_j-b_j$, we obtain the current
\begin{widetext}
\begin{align}\nonumber
 I_D &= \frac{eT}{2} \Bigl [ t_1^2 t_2^2 {\cal F}(\delta L) 
     + r_1^2 t_2^2 {\cal F}(\delta L - 2 d_1)
     + t_1^2 r_2^2 {\cal F}(\delta L - 2 d_2)
     +r_1^2 r_2^2 {\cal F} (\delta L +2 d_1 - 2 d_2)  
     - 2 t_1 t_2 r_1 r_2 {\cal F}(\delta L + d_1 - d_2)
     \nonumber\\     &\quad
 +2 t_1 t_2 r_1 r_2 {\cal F}(\delta L - d_1 - d_2) \Bigr],\quad\text{where}\quad
 t_j = \frac{1-\lambda_j^2}{1+\lambda_j^2},\quad
 r_j = \frac{2 \lambda_j}{1+\lambda_j^2},
\quad {\cal F}(X) = \frac{\sin(eVX/v)}{\sinh(\pi T X/v)}.
\label{IVnoEV}
\end{align}
\end{widetext}
In Eq.~\eqref{IVnoEV}, we assumed no bulk MZM defects inside the interferometer, as we will also assume in Sec.~\ref{sec3}. However, we will discuss their effects on transport in Sec.~\ref{sec4}. Equation \eqref{IVnoEV} describes electric transport from source to drain carried by the interference between a pair of Majorana fermions originating at the source and fused at the drain. At the line junction $j\in \{1,2\}$, a Majorana fermion has probability amplitude $t_j$ to remain in the same edge channel and amplitude $r_j$ to switch channels, where
$t_j^2+r_j^2=1$ holds by unitarity. 

Each term in Eq.~\eqref{IVnoEV} corresponds to a distinct process for the pair of Majorana fermions. The factor ${\cal F}(X)$ represents interference, where $X$ is the path-length difference; its numerator arises from dynamical phases at energy $eV$, while the denominator is due to the fact that quantum coherence is preserved only for $|X| \lesssim l_T$
with the thermal length $l_T = v/(\pi T)$. For example, the first term in Eq.~\eqref{IVnoEV} describes the process where both Majorana fermions  stay in their respective edge channel, where the probability is $t_1^2 t_2^2$ and the path-length difference is $X=\delta L$. The second term corresponds to a process where both Majorana fermions switch channels at the first junction, and hence the probability is $r_1^2 t_2^2$ and the path-length difference is $X = [b_1 + (L_1 - a_1)] - [a_1 +(L_2 -b_1) ]=\delta L -2 d_1$. Similar reasoning applies to the remaining terms.

\begin{figure}[tb] 
    \centering
    \includegraphics[width=\linewidth]{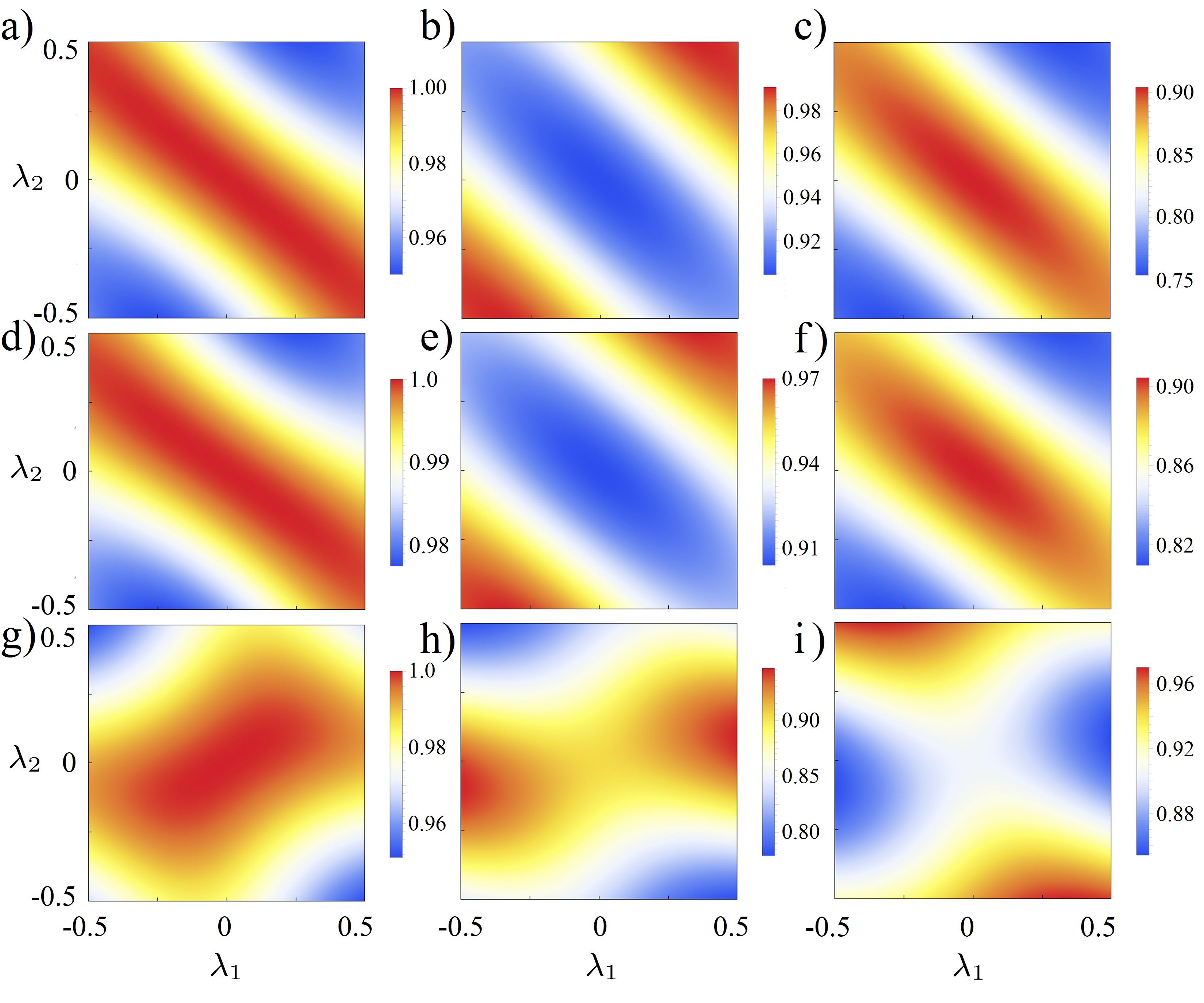} 
    \caption{2D color-scale plots for the finite-temperature
    DC conductance $G$ in units of $G_0=e^2/h$ at vanishing voltage ($V\to 0$) in the 
    $(\lambda_1,\lambda_2)$--plane for different asymmetric device configurations.  The shown results are obtained from Eq.~\eqref{IVnoEV} and hold in the absence of EV tunneling.  The interferometer arm lengths were chosen $L_1 = 0.5 v/T$ and $L_2 = L_1/2$ for $\delta L>0$, and swapped for $\delta L<0$. The tunneling locations  $\{a_1,b_1,a_2,b_2\}$ are   permutations of the values $\{0.035,0.08,0.11,0.175\}\,v/T$. Different panels correspond to either $\delta L=0$ [panels (a),(d), and (g)], $\delta L<0$ [panels (b),(e), and (h)], or $\delta L>0$ [panels
    (c),(f), and (i)].  The tunneling points are ordered as $a_1<a_2<b_1<b_2$ in panels (a),(b), and (c). The ordering $a_1<b_1<a_2<b_2$ applies to panels (d),(e), and (f). Finally, $a_1<b_1<b_2<a_2$ holds for panels (g),(h), and (i).
    For the symmetric device studied in Ref.~\cite{nava2024}, one simply has $G=G_0$ independent of $(\lambda_1,\lambda_2)$.  }
    \label{fig2}
\end{figure}

In the symmetric case, $\delta L=d_1=d_2=0$, Eq.~\eqref{IVnoEV} reduces to the familiar result
$I_D=\frac{e^2}{h}V$ \cite{fu2009,akhmerov2009}, where all Majorana fermion tunneling effects cancel out. We then obtain the DC conductance $G=G_0$ with the conductance quantum $G_0=e^2/h$. In fact, at asymptotically low temperatures and low voltages (such that $I_D\propto V$), where one can linearize all sine and sinh functions in Eq.~\eqref{IVnoEV} with ${\cal F}(X)\to eV/(\pi T)$, the result $G=G_0$ also holds for generic asymmetric setups due to unitarity.  
At finite temperature, however, the DC conductance in general depends on the Majorana tunneling amplitudes $\lambda_j$. This is illustrated in Fig.~\ref{fig2},
where we show how the length mismatch $\delta L$ and the ordering of the tunneling points $(d_j)$ affect the DC conductance and its structure in the $(\lambda_1,\lambda_2)$--plane.
We note in passing that as $\delta L$ decreases, the transition between the two regimes in Fig.~\ref{fig2}(a) and Fig.~\ref{fig2}(b) proceeds through a separation along the $\lambda_2=\lambda_1$ direction, followed by a counter-clockwise $\pi/2$--rotation in the $(\lambda_1,\lambda_2)$--plane.

\section{Edge vortex tunneling and chiral bosonization} \label{sec3}

In this section, we include EV tunneling processes at the line junctions in Fig.~\ref{fig1}.  In general, it is highly advantageous to employ the chiral bosonization approach to tackle the effects of EV tunneling  \cite{fendley2007,nava2024,giuliano2026}.   However, since for $L_1\ne L_2$, one cannot combine
the two Majorana fields $\xi_1(x)$ and $\xi_2(x)$ into a single Dirac fermion field which could then be bosonized,
the asymmetric device in Fig.~\ref{fig1} requires a more general approach than the one taken in Ref.~\cite{nava2024} for  $L_1=L_2$.  We introduce this approach in Sec.~\ref{sec3a}, 
followed by a discussion in Sec.~\ref{sec3b} on how to compute the frequency-dependent conductance using this framework.  Finally, we address EV tunneling processes in Sec.~\ref{sec3c}.

\subsection{Chiral bosonization in the doubling framework}\label{sec3a}

Our approach below is to employ \emph{two replicas} for each field,  $\xi_{j=1,2}(x)\to \xi_j^{(a)}(x)$ with replica index $a=1,2$, and similarly for $\Psi^{(a)}_S(x)$ and $\Psi_D^{(a)}(x)$. 
The algebra \eqref{MajoAlgebra} for the Majorana fermion fields then contains another Kronecker symbol $\delta_{a,a'}$, and
Eqs.~\eqref{MajoranaHam0} and \eqref{HSD} involve the replica index $a=1,2$. However, both replicas evolve completely independently under the full theory since the introduction of replicas is only a technical means used to implement bosonization.
For instance, in the doubled formalism, Majorana fermion tunneling is described by
\begin{equation}
 H_{\rm MT} \to - iv\sum_{j=1}^2 \lambda_j \sum_{a=1}^2 \, \xi^{(a)}_1 (a_j) \,\xi^{(a)}_{2} (b_j)  .
\label{MTfermR}
\end{equation}
Similarly, the matching conditions \eqref{match1} at $x=0$ and $\tilde x=0$ now hold separately for each replica index.
The doubling procedure then allows us to combine the two Majorana fields for the upper arm, $\xi_1^{(1)}(x)$ and $\xi_1^{(2)}(x)$, to a single Dirac field which in turn can be bosonized in terms of a chiral boson field $\phi_u(x)$; the index reminds us that this field refers to the upper edge.  Similarly, we can proceed for the lower edge, leading to the bosonization expressions \cite{delft1998,gogolin1998}
\begin{align}
 \xi^{(1)}_1(x) + i \xi_1^{(2)} (x)   &= \frac{2\eta_u}{\sqrt{2\pi l_c}} e^{i\phi_u (x)},\quad 0<x<L_1,  \\
      \xi^{(2)}_2(x) + i \xi_2^{(1)} (x) &= \frac{2\eta_d}{\sqrt{2\pi l_c}} e^{i\phi_d (x)},\quad 0<x<L_2, \nonumber 
\end{align}
where $\eta_{u,d}$ are Klein factors and $l_c=v/\Delta$ is a short-distance cutoff length. The chiral boson fields obey the commutator algebra ($\alpha,\alpha'\in\{u,d\}$)
\begin{equation}\label{bosonalgebra}
[ \phi_\alpha(x), \phi_{\alpha'}(x') ] = i\pi \delta_{\alpha,\alpha'}\,{\rm sgn}(x-x') .
\end{equation}
Similarly, we bosonize the replicated chiral Dirac modes using chiral boson fields 
$\Phi^{(a)}_S(x)$ for incoming particles with $-L_0<x<0$ and $\Phi^{(a)}_D(\tilde x)$
with $0<\tilde x<L_0$ for outgoing particles,
\begin{equation}
    \Psi^{(a)}_S(x)=\frac{\eta_{S}^{(a)}}{\sqrt{2\pi l_c}}e^{i\Phi^{(a)}_S(x)},
    \quad \Psi^{(a)}_D(\tilde x)=\frac{\eta_{D}^{(a)}}{\sqrt{2\pi l_c}}e^{i\Phi^{(a)}_D (\tilde x)},
\end{equation}
The algebra of these boson fields follows again from Eq.~\eqref{bosonalgebra}. 
All Klein factors above, collectively denoted by $\eta_k$, can be represented as Majorana fermions with $\eta_k^\dagger=\eta_k^{}$ and the anticommutator relation $\{\eta_k,\eta_{k'}\}=2\delta_{k,k'}$. Their role is to ensure anticommutation rules between different fermion types, while boson fields with the algebra \eqref{bosonalgebra} 
 establish anticommutation rules for a given fermion type.

 The bosonized expressions for the doubled versions of $H_0$, $H_S$, $H_D$, and $H_{\rm MT}$ are then given by 
\begin{eqnarray}\nonumber
    H_0 &=& \frac{v}{4\pi}  \int_0^{L_1} \! \! dx \, [\partial_x\phi_u(x)]^2 + \frac{v}{4\pi}\int_0^{L_2}\!\! dx\,[\partial_x\phi_d(x)]^2 ,\\ \nonumber 
    H_S &=& \frac{v}{4\pi} \int_{-L_0}^0 \!\! dx\, \sum_a [\partial_x\Phi^{(a)}_S(x)]^2, \\ \nonumber
     H_D &=&\frac{v}{4\pi} \int_{0}^{L_0} \!\! d\tilde x\, \sum_a [\partial_{\tilde x}\Phi^{(a)}_D(\tilde x)]^2,\\
H_{\rm MT} &=& - i\eta_u \eta_d \frac{2v}{\pi l_c} \sum_{j} \lambda_j  \sin [\phi_u (a_j)+\phi_d(b_j)].
\label{doubl.3x}
\end{eqnarray}
In the replicated bosonized version, the matching conditions \eqref{match1} take the form
\begin{eqnarray}\nonumber
\eta_{S}^{(1)} e^{i\Phi_S^{(1)}(0)} &=& \eta_u \cos [\phi_u (0)] + i \eta_d \sin [\phi_d (0)],   \\
\eta_{S}^{(2)} e^{i\Phi_S^{(2)} (0)} &=& \eta_u \sin [\phi_u (0)] + i \eta_d \cos [\phi_d (0)], \nonumber \\
\eta_{D}^{(1)} e^{i\Phi_D^{(1)} (0)} &=&\eta_u \cos [\phi_u (L_1)] + i \eta_d \sin [\phi_d (L_2)],\nonumber \\
\eta_{D}^{(2)} e^{i\Phi_D^{(2)} (0)} &=& \eta_u \sin [\phi_u (L_1)] + i \eta_d \cos [\phi_d (L_2)]. \label{dd.3}
\end{eqnarray}

\subsection{Transport quantities}\label{sec3b}

Being interested in transport properties, we include the (generally time-dependent) applied bias voltage $V(t)$ in the doubled bosonized theory through the term
\begin{equation}\label{HV}
    H_V (t) = - \frac{e}{2\pi}  \sum_{a} \int_{-L_0}^0 \!\! d x \, V(t)\, \partial_x \Phi_{S}^{(a)} (x) ,
\end{equation}
where we choose a gauge with chemical potential $\mu_D=0$ in the drain electrode.
In Eq.~\eqref{HV}, we used that $\partial_x \Phi_S(x)$ determines the local charge density fluctuations in the incoming chiral Dirac channel \cite{gogolin1998}. On the other hand, the current $I_D(t)$ reaching the drain electrode at time $t$ can be read out already at $\tilde x=0$ (up to an irrelevant time shift $L_0/v$), 
since all particles entering the outgoing chiral Dirac channel are absorbed by the drain electrode.  
For a given replica $a$, the corresponding operator is given by
\begin{equation}\label{currop}
   \hat I_D^{(a)} = \frac{ev}{2\pi} \partial_{\tilde x} \Phi_{D}^{(a)} (\tilde x=0) .
\end{equation}
Switching to frequency space and exploiting that both replicas are fully decoupled, the expectation values
$I_D^{(a)}(\omega)=\langle \hat I_D^{(a)}(\omega)\rangle\equiv I_D(\omega)$ must be independent of $a$.  
In the linear-response regime $eV\ll T$, one can then express the AC conductance $G(\omega)$ in terms of the equilibrium ($V=0$) retarded current-current correlation function $\Pi^R(t)$ through a Kubo formula \cite{altland2010,giuliano2026},
\begin{equation}\label{conddef}
    G(\omega)=\frac{I_D(\omega)}{V(\omega)} = \frac{\widetilde\Pi^R(\omega)}{i\omega},\quad
    \widetilde\Pi^R(\omega)=\int dt \, e^{i\omega t} \, \Pi^R(t),
\end{equation}
with  
\begin{equation}\label{pir}
 \Pi^R(t)= -i\Theta(t) \, \left\langle \left[\hat I_D^{(a)}(t),
\hat I^{(a)}_S(0)\right]\right \rangle,
\end{equation}
where $\hat I_S^{(a)}(t)=\frac{ev}{2\pi}\partial_x\Phi_S^{(a)}(x=0,t)$ follows as in Eq.~\eqref{currop}. The replica index $a$ can be freely chosen in Eq.~\eqref{pir}, and $\Theta$ denotes the Heaviside step function.
One can obtain $\Pi^R(t)$ from its imaginary-time version,
\begin{equation}\label{piimag}
\Pi (\tau ) = - \left\langle {\cal T}_\tau \hat I_{D}^{(a)} (\tau) \hat I_{S}^{(a)} (0) \right\rangle  ,
\end{equation}
where ${\cal T}_\tau$ is the imaginary-time ordering operator,
through a standard Wick rotation \cite{altland2010}.  
In particular,  in view of the strict independence of both replicas, by squaring Eq.~\eqref{piimag}, we  
obtain
\begin{equation} \label{ds.6}
    [\Pi(\tau)]^2 = \left\langle {\cal T}_\tau \hat I^{(1)}_{D} (\tau)  \hat I_{D}^{(2)} (\tau)\hat I_{S}^{(1)} (0) 
    \hat I_{S}^{(2)} (0) \right\rangle. 
\end{equation}
We next express Eq.~\eqref{ds.6} only in terms of the boson fields
$\phi_{u,d}(x,\tau)$ characterizing the chiral Majorana modes.  In fact, using the matching conditions \eqref{dd.3}, the current operators   in Eq.~\eqref{ds.6} can equivalently be expressed as
\begin{eqnarray}
\hat I_{D}^{(1)} (\tau) &=&  i\eta_u \eta_d\frac{ev}{\pi l_c} \cos [\phi_u (L_1,\tau)] \, \sin [\phi_d (L_2,\tau)],
\nonumber \\
\hat I_{D}^{(2)} (\tau) &=& i\eta_u \eta_d\frac{ev }{\pi l_c} \sin [\phi_u (L_1,\tau)] \, \cos[\phi_d (L_2,\tau)] ,
\nonumber \\
\hat I_{S}^{(1)} (\tau) &=& i\eta_u \eta_d\frac{ev}{\pi l_c} \cos [\phi_u (0,\tau)] \,\sin [\phi_d (0,\tau)], \nonumber \\
\hat I_S^{(2)} (\tau) &=& i \eta_u \eta_d\frac{ev}{\pi l_c} \sin [\phi_u (0,\tau)] \,\cos [\phi_d (0,\tau)]. 
\label{ddx.3}
\end{eqnarray}
The average in Eq.~\eqref{ds.6} is then taken over $H=H_0+H_{\rm MT}+H_{\rm EV}$,
where the EV tunneling Hamiltonian $H_{\rm EV}$ is discussed in Sec.~\ref{sec3c} below. 
$H$ only depends on the boson fields $\phi_{u,d}(x,\tau)$.

As a sanity check for our formalism, we first address the case without Majorana tunneling and without EV tunneling, i.e., $H=H_0$, and compute the corresponding DC conductance $G=G(\omega=0)$.
In that case, Eq.~\eqref{ds.6} can be directly evaluated by performing Gaussian averages over the boson fields, with the result 
\begin{equation}\label{freepi2}
    [\Pi(\tau)]^2= \frac{(eT/2)^4}{\sin^2[( v\tau  - iL_1)/l_T]  \sin^2[(v\tau - iL_2)/l_T]},
\end{equation}
where we again use the thermal length $l_T = v/(\pi T)$.
For the symmetric device studied in Ref.~\cite{nava2024}, $\Pi(\tau)$ is given by taking the positive branch of the square root of the right-hand side of Eq.~\eqref{freepi2}. We use this  rule for selecting the  branch of the complex square root function throughout. 
Using Eq.~\eqref{freepi2} for arbitrary $\delta L=L_1-L_2$, i.e., going beyond Ref.~\cite{nava2024}, taking the square root with the positive sign, performing the Fourier transformation to frequency space, and finally taking the $\omega\to 0$ limit, the DC conductance in Eq.~\eqref{conddef} follows as
\begin{equation}
 \frac{G(\omega\to 0)}{G_0} =  \frac{\delta L/l_T}{\sinh (\delta L/l_T)}. 
\end{equation}
This result agrees with the DC conductance computed from
the scattering approach in Sec.~\ref{sec2} for $\lambda_j=0$, see Eq.~\eqref{IVnoEV}. 
In particular, for $\delta L\to 0$ and/or $T\to 0$, we recover $G=G_0$ as for the symmetric device \cite{fu2009,akhmerov2009}.

In a similar way, we have checked that corrections to $G(0)$ at lowest nontrivial order in the Majorana tunneling amplitudes $\lambda_j$ as obtained from Eq.~\eqref{IVnoEV}
are correctly recovered by the doubled bosonization approach with the above square-root prescription. 

\subsection{Edge vortex tunneling}\label{sec3c}

We are now ready to include EV tunneling processes for the general geometry of Fig.~\ref{fig1}.  Using the doubled bosonization approach  in Sec.~\ref{sec3a}, combined with insights from Refs.~\cite{difrancesco1987,fendley2006,fendley2007}, we compute correlation functions of primary fields of the 
2D critical Ising model (which correspond to the $\sigma$ operators) through their relation to the critical Ashkin-Teller model which
can be accessed by bosonization.  

To begin, let us  review the construction of composite EV tunneling operators for the symmetric interferometer with $\delta L=d_1=d_2=0$,
see Refs.~\cite{fendley2007,nava2024}, where one introduces an
auxiliary isospin-$1/2$ operator ${\bf S}=(S^x,S^y,S^z)$   to 
ensure correct outcomes of anyon fusion processes.
In general, the Ising anyon fusion rule, $\sigma\times \sigma\sim I+\psi$, allows for a vacuum ($I$) and a  fermionic $(\psi)$ fusion channel \cite{nayak2008}.  However, as explained  in Ref.~\cite{giuliano2026}, the $\psi$ channel is available only in multi-terminal interferometer geometries.  In the Mach-Zehnder interferometers studied in this paper, the vacuum fusion channel ($I$) is the only option, since no 
degree of freedom is available to account for the associated parity change of neutral fermions.
In the symmetric geometry, tunneling of EVs between both arms at the specific location $x=a_j$ (with $b_j=a_j$) is then described by the 
operator  \cite{fendley2007,nava2024}
\begin{equation}
{\cal T}_\sigma (a_j) = \sigma_u (a_j) \sigma_d (a_j)   =  S^- e^{\frac{i}{2} \phi (a_j)} + S^+ e^{-\frac{i}{2} \phi (a_j)} , 
\label{vot.8}
\end{equation}
where $S^\pm=S^x\pm iS^y$ and the EV operators $\sigma_u$ ($\sigma_d$) refer to the upper (lower) arm. We here used the conventional (non-doubled) bosonization approach as appropriate for the symmetric device, i.e., when $\phi \equiv\phi_{u} = \phi_{d} $.  Since a quantum phase slip on the central SC island creates (or annihilates) EV pairs at both junctions simultaneously, 
the  composite EV tunneling Hamiltonian involves a product of four $\sigma$ operators, 
\begin{align}\label{vtt.2}
H_{\rm EV} &= \frac{\Gamma}{2} \{{\cal T}_\sigma (a_1),{\cal T}_\sigma (a_2) \} \\ \nonumber
&= \frac{\Gamma}{4} \biggl[ (1-2S^z) e^{-2\pi i n_g} e^{\frac{i}{2}\phi (a_1) } e^{-\frac{i}{2}\phi (a_2) }
+\\\nonumber &\quad+ (1+2 S^z ) e^{2\pi i n_g} e^{-\frac{i}{2}\phi (a_1) }e^{\frac{i}{2}\phi (a_2) }  \biggr] + {\rm H.c.}
\end{align}
Here $S^z=\pm 1/2$ labels the conserved global fermionic parity sector. In the second equality in Eq.~\eqref{vtt.2}, we have taken into account  the gate parameter $n_g$. The composite EV creation (or annihilation) rate $\Gamma$ is estimated as \cite{schoen1990,nava2024,giuliano2026}
\begin{equation}\label{Gamma}
    \Gamma \simeq \omega_p \left( \frac{E^\ast_J}{2E_C} \right)^{1/4} e^{-\sqrt{8E^\ast_J/E_C}},
\end{equation}
where $\omega_p=\sqrt{8E_J^\ast E_C}$ is the plasma frequency and $E_J^\ast=2E_J$ is the Josephson energy of the central SC island; we assume $\omega_p\gg \Delta$.  Note that for a given choice of $S^z$, 
we have two contributions in Eq.~\eqref{vtt.2} which correspond to counter-clockwise and clockwise propagation of EVs around the central floating SC island, respectively \cite{nava2024}.

Now we turn to the general asymmetric geometry, where we switch to the doubled bosonization framework introduced in Sec.~\ref{sec3a}.
Let us first consider the product of four $\sigma$ operators at the junction with $x=a_j$ on the upper
arm and $x=b_j$ on the lower arm (with fixed $j\in \{1,2\}$). The corresponding EV operators are denoted by $\sigma_{u}^{(a)}(a_j)$ for the upper arm, and by $\sigma_d^{(a)}(b_j)$ for the lower arm, with the replica index $a$. 
Motivated by our discussion in Sec.~\ref{sec3b} for the squared current-current correlation function, we consider the product of two pairs of operators with different replica indices, 
\begin{align} \nonumber
{\cal T}_\sigma(a_j,b_j)&= [\sigma_{u}^{(1)} (a_j) \sigma^{(1)}_{d} (b_j)] [\sigma^{(2)}_{u} (a_j) \sigma_{d}^{(2)} (b_j)]\\
&=  [\sigma_{u}^{(1)} (a_j) \sigma_{u}^{(2)}(a_j)] [\sigma_{d}^{(1)} (b_j) \sigma_{d}^{(2)} (b_j)]  \nonumber \\ \nonumber
&=  \left[ S_u^- e^{\frac{i}{2}\phi_u (a_j)}  +S_u^+ e^{-\frac{i}{2} \phi_u (a_j)}
  \right]  \\ &\quad\times  \left[  S_d^+ e^{ \frac{i}{2} \phi_d (b_j)}  + S_d^- e^{-\frac{i}{2} \phi_d (b_j)}   \right], 
\label{vot.9}
\end{align}
where we used the fact that $\sigma$ operators for different replicas commute (to obtain the second line) and employed Eq.~\eqref{vot.8} with auxiliary isospin-$1/2$ operators ${\bf S}_{u,d}$ (to obtain the third line).
Equation \eqref{vot.9} represents the doubled version of Eq.~\eqref{vot.8} for the asymmetric geometry.
The role of the isospin operators ${\bf S}_{u,d}$ is again to ensure that EV operators can only fuse to the vacuum channel. Terms like Eq.~\eqref{vot.9}  appear once $[\Pi(\tau)]^2$ in Eq.~\eqref{ds.6} is   expanded in powers of $\Gamma$.  At this stage, however, it 
is unclear whether ${\bf S}_u$ and ${\bf S}_d$ should be treated as independent
isospin variables, and one has to invoke physical arguments to make progress.  An important guideline is that the  calculations have to recover known results  \cite{nava2024} for the symmetric geometry.\\

\begin{figure} 
    \centering
    \includegraphics[width=\linewidth]{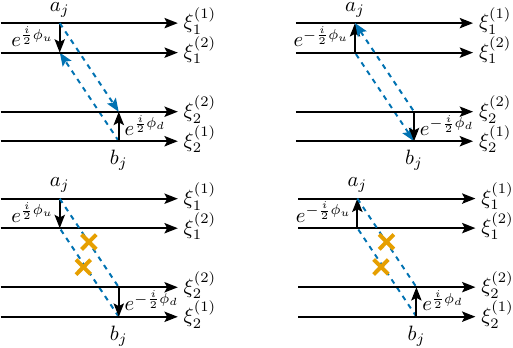} 
    \caption{EV tunneling processes  within the doubled theory, where we 
    show the Majorana fields $\xi_j^{(a)}(x)$ with replica index $a=1,2$. 
    The tunneling points are located at $x=a_j$ on the upper edge ($j=1$), and at $x=b_j$ on the lower edge ($j=2$).
    Bosonization implements vortex tunneling (black arrows) within a single edge while coupling the two replicas.
    For independent isospin operators ${\bf S}_{u,d}$, all four processes are allowed in Eq.~(\ref{vot.9}).
    However, not all of these correspond to physical processes within a single replica sector.
    While the pair of  Hermitian-conjugate processes shown in the top row corresponds to physical vortex tunneling (blue dashed arrows), the bottom row does not represent physical processes and  is therefore forbidden (orange crosses).
    This constraint is implemented by requiring that, when computing EV tunneling corrections to $[\Pi(\tau)]^2$, the total isospin $S^z_{u} + S^z_{d}$ is conserved. }
    \label{fig3}
\end{figure}

The corresponding lowest-order diagrams contributing to $[\Pi(\tau)]^2$ are schematically indicated in Fig.~\ref{fig3}.
Since we have doubled the system degrees of freedom, there are four 
possibilities as illustrated by the four panels in Fig.~\ref{fig3}. 
Out of the four processes, only two correspond to physical EV tunneling within a replica sector. 
As shown in Fig.~\ref{fig3}, the physical processes require  the total isospin $S^z_{u}+ S^z_{d}$ to be conserved because of global fermion parity conservation.
The operator ${\cal T}_\sigma(a_j,b_j)$ in Eq.~\eqref{vot.9} should then be replaced in the doubled theory by 
\begin{equation}\label{finalt}
  {\cal T}_\sigma(a_j,b_j) =
  S^- e^{\frac{i}{2} \phi_u (a_j)} e^{\frac{i}{2} \phi_d (b_j)} + S^+  e^{-\frac{i}{2} \phi_u (a_j)} e^{-\frac{i}{2} \phi_d (b_j)}.
\end{equation}
 By switching from the two isospin operators 
${\bf S}_{u,d}$ to a single isospin-$1/2$ operator ${\bf S}$, we single out the physically allowed diagrams. In expansions for $[\Pi(\tau)]^2$ in powers of $\Gamma$, one can then employ the following (doubled) EV Hamiltonian for composite EV tunneling processes,
\begin{widetext}
\begin{align}
 H_{\rm EV} = \frac{\Gamma}{2} \{{\cal T}_\sigma(a_1,b_1) , {\cal T}_\sigma(a_2,b_2) \}&= \frac{\Gamma}{4} \biggl[ (1-2 S^z ) e^{-2\pi i n_g}e^{\frac{i}{2} \phi_u (a_1)}
e^{\frac{i}{2} \phi_d (b_1)} e^{-\frac{i}{2} \phi_u (a_2) }
e^{-\frac{i}{2}\phi_d (b_2) }  \nonumber \\ &\quad+ (1+ 2 S^z ) e^{2\pi i n_g}e^{-\frac{i}{2} \phi_u (a_1)}e^{-\frac{i}{2} \phi_d (b_1)} 
e^{\frac{i}{2} \phi_u (a_2) } e^{\frac{i}{2}\phi_d (b_2) } \biggr] + {\rm H.c.} , \label{vvtt.2}
\end{align}
\end{widetext}
where we have again restored the $n_g$ dependence.  

\section{Calculation of transport observables} \label{sec4}

In this section, we employ the formalism developed in Sec.~\ref{sec3} in order to analyze the linear-response conductance corrections, $\delta G(\omega)$, arising from EV tunneling
as described by $H_{\rm EV}$ in Eq.~\eqref{vvtt.2} to first order in $\Gamma$.  In Sec.~\ref{sec4a}, we consider the case without Majorana fermion tunneling and show that there is no
conductance correction in the DC limit, $\delta G(0)=0$.  For finite $\omega$, the most interesting
contribution is a capacitive term encoded by an effective capacitance $C_{\rm eff}$.  While this quantity
has been characterized for the symmetric device in Ref.~\cite{nava2024}, we here provide results for the generic asymmetric geometry in Fig.~\ref{fig1}.  In Sec.~\ref{sec4b}, we then
discuss what happens if one simultaneously has Majorana fermion tunneling and EV tunneling processes.

\subsection{Conductance corrections due to EV tunneling}
\label{sec4a}

We start with the first-order contribution to  
$[\Pi(\tau)]^2$ in Eq.~\eqref{ds.6}, %\FH{Also here changed (1) to (2) in the second source current. I am wondering, if there are contributions of the form $\Pi^2 = \Pi^\text{MT} \delta \Pi$ where $\Pi^\text{MT}$ are Majorana tunneling. I guess not, but we might want to comment? }
\begin{multline}\label{dp2}
    [\delta \Pi(\tau)]^2= \int_0^{1/T} d\tau'\Bigl \langle {\cal T}_\tau
    \hat I^{(1)}_{D} (\tau)  \hat I_{D}^{(2)} (\tau)\hat I_{S}^{(1)} (0) 
    \hat I_{S}^{(2)} (0) \\
			 \times H_{\rm EV}(\tau') \Bigr\rangle  + {\cal O}(\Gamma^2),
\end{multline}
with the bosonized current operators in Eq.~\eqref{ddx.3} and $H_{\rm EV}$ in Eq.~\eqref{vvtt.2}.  We here consider the situation without Majorana fermion tunneling, i.e., $\lambda_j=0$.
The average in Eq.~\eqref{dp2} is thus taken with respect to the Gaussian model $H_0$ in Eq.~\eqref{doubl.3x}. 

\begin{figure}[t] 
    \centering
    \includegraphics[width=0.7\linewidth]{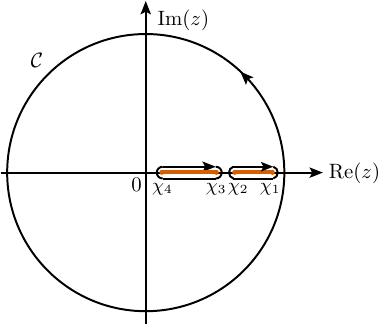} 
    \caption{Complex integration contour ${\cal C}$ and branch cuts for the evaluation of the complex integral in Eq.~\eqref{vox.11}. On the positive real axis, two branch cuts with ${\rm Im}(z)=0$ and
     $\chi_4<{\rm Re}(z)<\chi_3$ and $\chi_2<{\rm Re}(z)<\chi_1$ exist, see Eq.~\eqref{chij}.  
     The positions of the branch points $\chi_j$ correspond (via the mapping $x\mapsto\exp(-2\pi Tx/v)$) to the location of the EVs (all four EVs are evaluated at teh same imaginary time). The branch cuts reflect the Ising anyon nature of the EV. In the symmetric case $a_j =b_j$, the branch cuts disappear and a bosonization within a single replica sector becomes possible.
 } \label{fig4}
\end{figure}

After tedious but straightforward calculations, using the square root prescription discussed in Sec.~\ref{sec3c} and switching to bosonic Matsubara frequencies $\omega_n>0$, we obtain 
\begin{eqnarray}\nonumber
  &&  \widetilde{\delta \Pi}(\omega_n)=\frac{\Gamma \cos\left(4\pi S^z s_\sigma+\alpha\right) (eT/4)^2}
     {\left[ \Bigl(\frac{l_T}{l_c}\Bigr)^2 \sinh\left(\frac{a_2-a_1}{l_T}\right) \sinh\left(\frac{b_2-b_1}{l_T}\right)\right]^{1/8}}\times\\ 
  &&\times    {\cal A}\left(\{a_j,b_j\};\omega_n\right)
    {\cal A}\left(\{L_1-a_j,L_2-b_j\};\omega_n\right),   \label{deltapi}
\end{eqnarray}
where $l_c=v/\Delta$ and the specific ordering $a_1\le b_1<a_2\le b_2$ of the tunneling points has been assumed.  (However, the calculation can be carried out for any other physically allowed ordering.)  The term with $s_\sigma=1/16$ 
 in Eq.~\eqref{deltapi} arises from the bosonic algebra \eqref{bosonalgebra} and is a signature of the topological spin \cite{nava2024}, see Eq.~\eqref{correl}, which in turn encodes the non-Abelian braiding statistics of Ising anyons.  The cosine contains the offset
 angle $\alpha=2\pi n_g$ encoding the dependence on the gate charge parameter $n_g$.  In the presence of Majorana tunneling, this offset angle receives additional contributions from $\lambda_j$ terms, see Ref.~\cite{nava2024}.
The quantity ${\cal A}(\{a_j,b_j\};\omega_n)$   in Eq.~\eqref{deltapi} can be written in complex-plane integral representation as
\begin{eqnarray}\nonumber
&& {\cal A}\left(\{a_j,b_j\};\omega_n\right) = 
-\frac{2 l_T }{v }  (\chi_1\chi_2\chi_3\chi_4)^{1/4} \times \\ &&\times 
\left[ \sinh\left(\frac{a_2-a_1}{l_T }\right)\sinh \left( \frac{b_2-b_1}{l_T}\right) \right]^{1/2} \times
\label{vox.11}\\   
&& \times\oint_{\cal C} d z \: 
\frac{z^n}{\prod_{i=1}^4\sqrt{z-\chi_i}} \nonumber 
, 
\end{eqnarray}
where ${\cal C}$ is the unit circle in counter-clockwise orientation, see Fig.~\ref{fig4}. The natural number $n$ in Eq.~\eqref{vox.11} follows from $\omega_n=2\pi nT>0$, and we define the four real numbers $\chi_j$, $j=1,\dots, 4$ to be the sorted rearrangement of 
\begin{align}\label{chij}
    \{ e^{-2  a_1/l_T}, e^{-2  b_1 / l_T},
    e^{-2 a_2/l_T}, e^{-2  b_2/l_T} \}\,, 
\end{align}
with $0<\chi_4\le \chi_3<\chi_2\le \chi_1<1$.
Branch cuts for the complex integrand in Eq.~\eqref{vox.11} are then located on the real axis 
between $\chi_4$ and $\chi_3$, and between $\chi_2$ and $\chi_1$, see Fig.~\ref{fig4}.
Note that for $a_j=b_j$, the branch cuts disappear, which is why bosonization within a single replica sector becomes possible in the symmetric case.
Similarly, ${\cal A}\left(\{L_1-a_j,L_2-b_j\};\omega_n\right)$ can be reduced to an integral in which $\{\chi_j\}$ are replaced by $\{\tilde \chi_j\}$, which  represent the sorted rearrangement of
\begin{multline}\label{tildechij}
    \{ e^{-2 (L_1-a_1)/l_T}, e^{-2  (L_2-b_1) / l_T},\\
    e^{-2  (L_1-a_2)/l_T}, e^{-2  (L_2-b_2)/l_T} \}. 
\end{multline}

Rotating back to real frequencies, $i\omega_n\to \omega+i0^+$ with ${\cal A}(\omega_n)\to A(\omega)$, and expanding in powers of $\omega$ around $\omega = 0$, Eq.~\eqref{vox.11} yields 
\begin{eqnarray}\nonumber
 && A\left(\{a_j,b_j\}; \omega\right) = -  \frac{2\omega l_T^2}{v^2}
 (\chi_1\chi_2\chi_3\chi_4)^{1/4} \, Y\left(\{a_j,b_j\}\right)  \\ &&\times
 \sqrt{ \sinh\left(\frac{a_2-a_1}{l_T }\right)\sinh \left(\frac{b_2-b_1}{l_T}\right)} \label{vv.5prequel}
  + {\cal O}(\omega^2),
\end{eqnarray}
with the dimensionless function
\begin{eqnarray}\label{vv.6bis} 
Y\left(\{a_j,b_j\}\right) &=& 
 \int_{\chi_4}^{\chi_3} \!\! d x  \frac{\ln x }{\sqrt{\prod_{k=1}^4 |x-\chi_k|}} \\  \nonumber
&-& \int_{\chi_2}^{\chi_1} \!\! d x \frac{\ln x}{\sqrt{\prod_{k=1}^4 |x-\chi_k|}} .
\end{eqnarray}
To arrive at Eq.~\eqref{vv.5prequel}, we have used Cauchy's theorem to evaluate 
the complex integral in Eq.~\eqref{vox.11} with the integration path shown in Fig.~\ref{fig4}.  

Since one similarly finds $A\left(\{L_1-a_j,L_2-b_j\};\omega\right)\propto \omega$ to leading order in $\omega$, inserting those results for the low-frequency behavior of the $A$-functions into (the analytic continuation to real frequencies of) Eq.~\eqref{deltapi}, 
we find $\widetilde{\delta \Pi}(\omega)\propto \omega^2$. As a 
consequence, Eq.~\eqref{conddef} implies
that there are \emph{no corrections to the DC conductance} due to EV tunneling and thus no signatures of the non-Abelian statistics.  This generalizes the corresponding result for the symmetric geometry \cite{nava2024} to the general geometry in Fig.~\ref{fig1}.

The leading low-frequency corrections to the AC conductance are  
linear in frequency and can be attributed to an effective capacitance
$C_{\rm eff}$. Allowing for $n_v$ localized MZMs inside the central SC region \cite{fendley2007}, the low-frequency AC conductance correction due to EV tunneling processes in the absence of Majorana tunneling (i.e., for $\lambda_j=0$)
 takes the form
\begin{equation}\label{ACcond}
    \delta G(\omega) = -i (-1)^{n_v} \omega C_{\rm eff}.
\end{equation}
Using the above expressions, quantitative results for $C_{\rm eff}$ 
for the asymmetric geometry in Fig.~\ref{fig1} follow.
We obtain 
\begin{eqnarray}\nonumber
C_{\rm eff}  &=&  \Gamma \cos \left(\frac{\pi}{4} S^z +\alpha \right) 
\frac{e^2 e^{-(L_1+L_2)/l_T }}{4 \pi^4 T^2}   (l_c/l_T)^{1/4} \\ \nonumber
&\times& \left[\sinh \left(\frac{a_2-a_1}{l_T}\right)  \sinh \left(\frac{b_2-b_1}{l_T}\right) \right]^{7/8} \\
&\times& Y\left( \{a_j,b_j\} \right)  \,
Y\left( \{L_1-a_j,L_2-b_j\}\right). \label{Ceff}
\end{eqnarray}
This general result for the effective capacitance consists of several factors which encode different signatures of the non-Abelian nature of the EVs.
As discussed in Ref.~\cite{nava2024}, the topological spin enters through the phase offset $\pi S^z/4 = \pm 2\pi s_\sigma$, while the conformal dimension $h_\sigma=1/16$ fixes, for example, the exponent of the cut-off dependence $\propto l_c^{4 h_\sigma}$ of $C_{\rm eff}$.
Moreover, the factor $Y( \{a_j,b_j\} )$ of Eq.~\eqref{Ceff} corresponds to the amplitude of splitting the vacuum at the source ($x=0$) into four EVs at the positions $a_1, b_1, a_2, b_2$.
Similarly, the factor $Y( \{L_1-a_j,L_2-b_j\} )$   describes the fusion process, which in turn depends on the distances $L_1 -a_j$ and $L_2-b_j$ of the EVs to the drain.

Let us now discuss a limiting case for the general result \eqref{Ceff} where analytical progress can be made.  We set $\delta L=0$ but put $b_j=a_j+\varepsilon_j$ with small distances $\varepsilon_{1,2}$ that serve as expansion parameters.  To lowest nontrivial order in $|\varepsilon_j|$, we find from Eq.~\eqref{Ceff}
\begin{eqnarray}
    \label{Ceffunequal}
   C_{\rm eff} &=& \Gamma \frac{e^2(a_2-a_1)(b_2-b_1)}{4 v^2}\cos\left(\frac{\pi}{4} S^z+\alpha \right)  \\ 
   &\times&\, \left[ \frac{l_T}{l_c} \sinh\left(\frac{a_2-a_1}{l_T}\right) \right]^{-1/4} + {\cal O}(\varepsilon_j^2).	\nonumber 
\end{eqnarray}
 For $\varepsilon_j\to 0$ and $b_j= a_j$, we reproduce the result of Ref.~\cite{nava2024}.
 We note that for $\delta L\ne 0$, finite corrections arise to order $\delta L^2$.

\begin{figure}[t] 
    \centering
    \includegraphics[width=\linewidth]{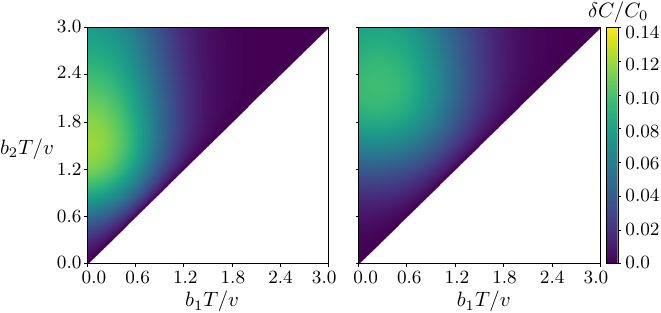}
    \caption{2D color-scale plot of the amplitude $\delta C$ of the effective capacitance $C_{\rm eff}$ [in units of $C_0=\Gamma e^2 (l_c T /v)^{1/4}/T^2$] as a function of the edge-vortex positions  $b_1$ and $b_2$ in the bottom edge mode, using $a_1 = 0.5 \,v/T$ and $a_2 = v/T$.  The interferometer arm lengths are set to $L_1=L_2 =3 v/T$  (left panel) and $L_1 = 3v/T$, $L_2 =4v/T$ (right panel). Note that $C_{\rm eff}$ oscillates as a function of the offset angle $\alpha$ as $\cos( \pi S^z/4 + \alpha)$. The phase offset $\pi S^z/4 = \pm 2\pi s_\sigma$ corresponds to the topological spin of EVs \cite{nava2024}.} \label{fig5}
\end{figure}

The general behavior of the maximal amplitude $\delta C$ of the effective capacitance $C_{\rm eff}$ is shown in Fig.~\ref{fig5}, as obtained from numerical evaluation of Eq.~\eqref{Ceff}.
The left panel illustrates the case of equal-length arms, $L_1 = L_2$. Remarkably, for fixed positions $a_1=0.5\,v/T$ and $a_2=v/T$ on the upper arm, the largest signal is \emph{not} obtained at the fully symmetric point $b_j=a_j$ but for a configuration in which the two EVs on the lower edge are separated further than those on the upper edge.
Over the range shown, the signal grows monotonically as $b_1\to 0$ and is maximal near $b_2\simeq 1.5\, v/T$.
This displacement follows from the structure of Eq.~\eqref{Ceff}. The factor $[\sinh (\frac{a_2-a_1}{l_T})  \sinh (\frac{b_2-b_1}{l_T})]^{7/8}$ grows with the separation $b_2-b_1$, whereas the splitting and fusion amplitudes $Y(\{a_j,b_j\})$ and $Y(\{L_1-a_j,L_2-b_j\})$ are suppressed once the mismatch $|b_j-a_j|$ exceeds the thermal length $l_T$. As a consequence, the optimal values  are determined by the competition between these two mechanisms.
The displacement is then itself of order $l_T$, and at the symmetric point the signal still amounts to about one half of its maximal value.
Importantly, we find that the signal remains large as long as all asymmetries stay within the range of the thermal length $l_T$.
Thus, the relevant length scale for EV interference is the thermal length, analogous to both the case of Majorana fermion interference discussed in Sec.~\ref{sec2} and the interference of quasiparticles in conventional quantum Hall edge states \cite{seelig2001,roulleau2008}.
We note that the size of the EV---corresponding to the injection time, as discussed in Ref.~\cite{nava2024}---is not considered here (essentially taken to be zero).
The estimated size of the EV was $\approx 0.1\,\mu$m in Ref.~\cite{nava2024}, while  $l_T \approx 1\,\mu$m at low temperatures ($T = 20\,$mK), confirming that, indeed, $l_T$ is the relevant scale.
Similarly, even an asymmetry in the total length ($\delta L\neq 0$) of the order of the thermal length  $l_T$ still allows essentially the full signal to be observed.
The right panel of Fig.~\ref{fig5} shows the case with $L_2 - L_1 = \pi l_T$.
Here, the optimal setup shifts to larger values of $b_2$, as the last factor in Eq.~\eqref{Ceff}, which corresponds to the fusion of the EVs, is optimized for $L_2 - b_2 \approx L_1 -a_2$.

\subsection{Conductance corrections with Majorana and EV tunneling}
\label{sec4b}

In the simultaneous presence of Majorana tunneling and EV tunneling, the conductance can also be evaluated using the doubled bosonization framework introduced in Sec.~\ref{sec3}. 
However, since the corresponding calculations are rather lengthy and eventually give little new insight, we here offer only a few qualitative remarks on the interplay of Majorana tunneling and EV tunneling in the asymmetric two-arm device in Fig.~\ref{fig1}.

First, for the AC conductance corrections $\delta G(\omega\ne 0)$ at low frequencies, the main effect of Majorana tunneling is captured by a shift of the phase $\alpha$ appearing in the effective capacitance in Eq.~\eqref{Ceff}.  Indeed, for the
symmetric device, one finds $\alpha=2\pi n_g-\pi(\lambda_1+\lambda_2)$ \cite{nava2024}.  For weak amplitudes $\lambda_j$ and
$\Gamma$, such a phase shift will be the leading effect of Majorana tunneling processes on the AC conductance.

One may now ask if the simultaneous presence  of Majorana tunneling and EV tunneling processes can allow for finite corrections to the DC conductance, $\delta G(0)$.  For the 
symmetric case, this is known \emph{not}\ to occur \cite{nava2024}, while for a multi-terminal geometry, the simultaneous presence of both tunneling mechanisms is even necessary for finite components of the DC conductance tensor \cite{giuliano2026}.  Now, for the general asymmetric two-arm interferometer in Fig.~\ref{fig1}, as shown in Sec.~\ref{sec2}, see in particular Fig.~\ref{fig2}, $\delta G(0)\ne 0$ even
for $\Gamma=0$.  Since EV tunneling gives at most small perturbative corrections on top of those results, we refrain from showing explicit calculations here. 

\section{Conclusions}\label{sec5}

We  generalized the transport theory of co-propagating Majorana interferometers to an asymmetric setup. 
To this end, we have developed a versatile bosonization method that  introduces a second fermionic replica.
We calculated the DC conductance due to finite Majorana tunneling and showed, akin to Ref.~\cite{nava2024}, that EV tunneling does not produce a correction to the DC conductance but only yields an effective capacitance $C_\text{eff}$ at low frequencies.
The capacitance exhibits oscillatory behavior as a function of the Majorana tunneling strength and gate charge $n_g$, with an offset due to the topological spin \cite{nava2024}.
Because this effect depends on EV interference, the arm lengths of the interfering paths must be comparable. We demonstrated that length asymmetries of the order of the thermal length $l_T =v/(\pi T)$ are permitted without substantially suppressing the interference. For example, with $l_T =1\,\mu$m at 20$\,$mK, this leads to manageable requirements for an experimental setup. We note that path-length differences can also arise from a velocity $v(x)$ that depends on position $x$; since the interference depends only on transit time, our results remain valid by replacing $x/v$ with $\int_0^x\!dx'/v(x')$.
Our findings open the possibility to study more complicated setups such as multi-terminal devices \cite{giuliano2026} or braiding of EVs \cite{beenakker2019b,adagideli2020}.

\section*{Data availability}

The data underlying the figures in this work will be made available at Zenodo \cite{Zenodo}.

\begin{acknowledgments}
We thank A. Akhmerov and C. Beenakker for discussions. We acknowledge funding by the Deutsche Forschungsgemeinschaft (DFG, German Research Foundation), Projektnummer 277101999 -- TRR 183 (project C01), under project No.~EG 96/15-1, and under Germany's Excellence Strategy -- Cluster of Excellence Matter and Light for Quantum Computing (ML4Q) EXC 2004/2 -- 390534769.  
\end{acknowledgments}

\bibliography{refs}

\end{document}